\documentclass[PAPER, atlasdraft=false, texlive=2023, cernpreprint, USenglish, texmf, orcidlogo]{atlasdoc}

\usepackage{atlaspackage}
\usepackage{atlasbiblatex}

\usepackage{atlasphysics}

\graphicspath{{logos/}{figures/}}
 
\usepackage{ANA-EXOT-2022-07-PAPER-defs}

\AtlasTitle{
Search for new phenomena in two-body invariant mass distributions using unsupervised machine learning for anomaly detection at $\sqrt{s} = 13$ TeV with the ATLAS detector}

\AtlasAbstract{
Searches for new resonances are performed using an unsupervised anomaly-detection technique.
Events with at least one electron or muon are selected from 140~fb$^{-1}$ of $pp$ collisions at $\sqrt{s} = 13$~TeV recorded by ATLAS at the Large Hadron Collider.
The approach involves training an autoencoder on data, and subsequently defining anomalous regions based on the reconstruction loss of the decoder.
Studies focus on nine invariant mass spectra that contain pairs of objects consisting of one light jet or $b$-jet and either one lepton ($e$, $\mu$), photon, or second light jet or $b$-jet in the anomalous regions.
No significant deviations from the background hypotheses are observed.
Limits on contributions from generic Gaussian signals with various widths of the resonance mass are obtained for nine invariant masses in the anomalous regions.
}

\author{The ATLAS Collaboration}

\AtlasRefCode{EXOT-2022-07}
 
\PreprintIdNumber{CERN-EP-2023-112}

\arXivId{2307.01612}

\AtlasJournalRef{\PRL 132 (2024) 081801}
\AtlasDOI{10.1103/PhysRevLett.132.081801}

\hypersetup{pdftitle={ATLAS document},pdfauthor={The ATLAS Collaboration}}
 
\begin{document}
 
\maketitle

\clearpage

\def\modelone{(1) charged Higgs boson production in association with a top quark, $tbH^{+}$ with $H^+\to t\bar{b}$}
\def\modeltwo{(2) a Kaluza--Klein gauge boson, $W_\text{KK}$, with the SM $W$ boson and a radion $\phi$}
\def\modelthree{(3) a $Z'$ boson decaying to a composite lepton $E$ and $\ell$, with $E \to Z \ell$}
\def\modelfour{(4) the SSM $W'\to W Z' \to \ell\nu q\bar{q}$}
\def\modelfive{(5) a simplified dark-matter model with an axial-vector mediator $Z' \to q\bar{q}$, where one of the quarks radiates a $W$ boson decaying to $\ell\nu$}
 
\def\fivemodel{\modelone;
\modeltwo;
\modelthree;
\modelfour;
\modelfive.
}
 
\def\fivemodelwithmass{\modelone;
\modeltwo;
\modelthree\ with a mass of 0.5~\TeV;
\modelfour;
\modelfive.
}
 
\def\fivemodelwithallmass{\modelone, with the $H^+$ mass of 2~\TeV;
(2) a 6~\TeV\ Kaluza--Klein gauge boson, $W_\text{KK}$, with the SM $W$ boson and a radion $\phi$;
(3) a 4~\TeV\ $Z'$ boson decaying to a composite lepton $E$ with a mass of 0.5~\TeV;
\modelfour, with the $W'$ mass of 6.2~\TeV\ and the $Z'$ mass of 6~\TeV;
\modelfive\ with a mass of 6~\TeV.
}
 
Searches for new physics phenomena beyond those described by the Standard Model (SM) require advanced techniques to devise selections that involve a large number of variables characterizing collision events.
Furthermore, limited understanding of how new physics would manifest itself has inspired the design of model-independent searches~\cite{EXOT-2016-38}.
In traditional methods, event selections are optimized to target specific signatures of signals beyond the SM (BSM signals) and to maximize their separation from SM background processes. Alternatively, event selection criteria can be relaxed to target more general signatures, but this reduces the ability to suppress background.
 
Machine learning (ML) anomaly-detection methods~\cite{Collins:2018epr,DAgnolo:2018cun,DeSimone:2018efk,Hajer:2018kqm,Heimel:2018mkt,Cerri:2018anq,Farina:2018fyg,Blance:2019ibf,Dillon:2020quc,Aarrestad:2021oeb,kasieczka2021lhc} provide a new way to study collision events.
One such approach uses an autoencoder (AE)~\cite{rumelhart1986learning,bourlard1988auto,hinton1994autoencoders,hinton2006reducing}, a neural network architecture that is commonly used in unsupervised learning.
The AE is trained using mostly SM background events and is applied to identify collision events that display kinematic properties different from those of SM events.
ATLAS previously used a weakly supervised learning technique for massive dijet final states~\cite{HDBS-2018-59} and an unsupervised machine-learning method to identify anomalous jets in a search for BSM resonances decaying into a Higgs boson and a generic new boson~\cite{ATLAS-CONF-2022-045}.
 
This Letter presents a generic search for resonances in various two-body final states that applies an anomaly detection method to the event topology, for the first time in ATLAS.
Events are triggered by the presence of an isolated lepton to reduce
contamination from QCD multijet events.
The two-body final states consist of jet\;\!+\;\!$Y$, where the jet can be a light jet or a $b$-jet (containing a $b$-hadron decay) while $Y$ can be a lepton (electron or muon), a photon, or another light jet or $b$-jet.
The highest-\pT candidate for each kind of final-state object is used, except for final states with either two light jets or two $b$-jets, where both the leading and subleading light jets or $b$-jets are considered.
Triggering on the lepton allows studies of invariant mass distributions below $1$~\TeV, which would be difficult to model because of trigger threshold effects if jet triggers were used~\cite{EXOT-2019-03}.
 
The data correspond to \SI{140}{\ifb} of $pp$ collisions at $\sqrt{s} = \SI{13}{\TeV}$ collected by the ATLAS experiment at the Large Hadron Collider (LHC).
The ATLAS detector~\cite{PERF-2007-01} is a multipurpose particle detector with cylindrical geometry.\footnote{ATLAS uses a right-handed coordinate system with its origin at the nominal interaction point in the center of the detector and the $z$-axis along the beam pipe.
The $x$-axis points to the center of the LHC ring, and the $y$-axis points upward. Cylindrical coordinates $(r, \phi)$ are used in the transverse plane, $\phi$ being the azimuthal angle around the $z$-axis. The pseudorapidity is defined in terms of the polar angle $\theta$ as $\eta = - \ln \tan (\theta/2)$.
The distance between two objects in $\eta$--$\phi$ space is $\Delta R = \sqrt{(\Delta \eta)^2 + (\Delta \phi)^2}$.
Transverse momentum is defined as $\pt = p \cdot \sin \theta$.}
It consists of an inner tracking detector (ID) surrounded by a superconducting solenoid, electromagnetic (EM) and hadronic sampling calorimeters, and a muon spectrometer with three toroidal superconducting magnets, providing a near \(4\pi\) coverage in solid angle.
A two-level trigger system is used to select events for storage.
Events used in this analysis were selected online by single-electron or single-muon triggers~\cite{TRIG-2016-01,TRIG-2018-05,ATL-DAQ-PUB-2019-001}.
An extensive software suite~\cite{ATL-SOFT-PUB-2021-001} is used in data simulation, in the reconstruction and analysis of real and simulated data, in detector operations, and in the trigger and data acquisition systems of the experiment.
 
The interaction vertex with the highest sum of the squared transverse momenta ($\pt^2$) of all associated tracks with $\pt > \SI{500}{\MeV}$ is selected as the primary vertex~\cite{ATL-PHYS-PUB-2015-026}. Electrons are reconstructed from energy clusters in the EM calorimeter that match a reconstructed track~\cite{EGAM-2018-01}.
Muons are reconstructed by combining a track from the ID with one from the muon spectrometer~\cite{MUON-2018-03}.
Electrons (muons) must have $E_{\text{T}} > \SI{20}{\GeV}$ ($\pt > \SI{20}{\GeV}$) and pseudorapidity $|\eta| < 2.47$ ($|\eta| < 2.5$), excluding the region $1.37 < |\eta| < 1.52$ for electrons.
To ensure that selected leptons originate from the primary vertex, their tracks must have $|d_0/\sigma(d_0)| < 5~(3)$ for electrons (muons) and $|\zzsth| < \SI{0.5}{\mm}$ for both lepton flavors.
Here $d_0$ and $\sigma(d_0)$ are the transverse impact parameter and its uncertainty, and $z_0$ is the longitudinal impact parameter along the beam line.
Electrons must satisfy the \enquote{Tight} likelihood-based identification criterion defined in Ref.~\cite{EGAM-2018-01}, and the \enquote{FCTight} isolation requirement defined in Ref.~\cite{TRIG-2018-05}.
Muons must satisfy the \enquote{Medium} cut-based identification criterion and the particle-flow-based \enquote{PflowTight} isolation requirement~\cite{MUON-2018-03}.
Photons are reconstructed from energy deposits in the central EM calorimeter~\cite{EGAM-2018-01}.
They must have $\pt > \SI{20}{\GeV}$ and pseudorapidity $|\eta| < 2.37$, excluding the region $1.37 < |\eta| < 1.52$.
The \enquote{tight} identification requirement and the \enquote{tight} isolation requirement are applied, both defined in Ref.~\cite{EGAM-2018-01}.
Jets are reconstructed using the anti-\kt algorithm~\cite{Cacciari:2008gp,Fastjet} with a radius parameter of $0.4$, applied to tracks in the ID and topological clusters~\cite{PERF-2014-07} processed using a particle-flow algorithm~\cite{PERF-2015-09}.
To reduce the effect of additional collisions in the same and neighboring bunch crossings (\pileup), jets with $\pt < \SI{60}{\GeV}$ and $|\eta| < 2.4$ must pass a jet-vertex-tagger requirement that has a \SI{96}{\%} selection efficiency for hard-scattered jets~\cite{PERF-2014-03}.
Jets containing a $b$-hadron are identified using the \SI{77}{\%} efficiency working point of the DL1r $b$-tagging algorithm~\cite{FTAG-2019-07}.
Jets failing to meet the $b$-tagging criteria are identified as light jets.
An overlap-removal procedure detailed in Ref.~\cite{TOPQ-2017-14} is applied to the selected objects.
 
The missing transverse energy, $\met$, is calculated from the vector sum of the transverse momenta of all reconstructed objects associated with the primary vertex.
To account for soft hadronic activity, a term including tracks associated with the primary vertex, but not with any of the reconstructed objects, is included in the $\met$ calculation~\cite{ATLAS-CONF-2018-023}.
 
Following the reconstruction of these objects, events are first selected by requiring at least one lepton ($e$ or $\mu$, denoted by $\ell$) with $\pt^{\ell} > \SI{60}{\GeV}$, and at least one jet with $\pt > \SI{30}{\GeV}$.
A trigger matching requirement~\cite{TRIG-2016-01} is applied to ensure that the reconstructed lepton lies in the vicinity of the corresponding trigger-level object.
Jets and $b$-jets are required to have a pseudorapidity of $|\eta| < \num{2.4}$.
These requirements constitute the event \textit{preselection}.
Nine invariant mass distributions are studied in this analysis.
The invariant mass distributions of \mjj and \mbb are reconstructed from the leading and sub-leading ($b$-)jets in each event; \mjb is reconstructed from the leading jet and leading $b$-jet; and \mje (\mbe), \mjm (\mbm), and \mjg (\mbg) are reconstructed from the leading ($b$-)jet and leading electron, muon, or photon, respectively.
An event that contains multiple types of objects can contribute to multiple mass distributions.
 
Several Monte Carlo (MC) samples were simulated to validate the analysis procedure, although the anomaly detection relies only on data samples.
All simulated SM samples, which are used to model the background, include a detailed ATLAS detector simulation~\cite{SOFT-2010-01} based on \GEANT~\cite{Agostinelli:2002hh}.
Samples for the benchmark BSM signal models instead include a faster, less detailed detector simulation~\cite{SOFT-2010-01}.
Additional $pp$ collisions generated using \Pythia~\cite{Sjostrand:2014zea} with the A3 set of tuned parameters~\cite{ATL-PHYS-PUB-2016-017} were overlaid to simulate the effects of \pileup interactions that match the multiplicity of additional collisions in data.
 
Kinematic features of the final-state objects in the preselected events are structured in a matrix called the rapidity--mass matrix (RMM) which is proposed as an input for machine learning~\cite{CHEKANOV201992}.
The RMM was tested for an anomaly detection method using MC event generators and was found to produce more robust AE training than when four-momenta are used as input~\cite{Chekanov:2021pus}.
The RMM is a square matrix employed to represent a comprehensive picture of the event in terms of popular variables used for BSM searches, such as $\met$, transverse energies, transverse masses, Lorentz factors, two-particle invariant masses, and two-particle rapidity differences.
In this analysis, the reconstructed final-state objects are light jets, $b$-jets, muons, electrons, or photons.
A maximum of ten light jets or $b$-jets are considered, along with up to five electrons, muons, and photons. Together with $\met$, a total number of 36 final-state objects are used to define the RMM.
The objects within each type are ordered in decreasing transverse energy. If the number of available objects for a particular type is less than the maximum allowed, the remaining elements in the corresponding row and column of the RMM are filled with zeros.
This ensures that the input size is the same for every event.

By construction, all elements of the RMM are defined to be between 0 and 1, and most variables are Lorentz-invariant under boosts along the longitudinal axis.
To reduce biases in the shapes of the jet\;\!+\;\!$Y$ invariant mass spectra, the nine invariant mass variables are excluded from the RMM.
The resulting input dimension is $36^2 - 9 = 1287$.
The RMM matrix is then flattened to a one-dimensional input vector before being fed into the AE.
 
The AE is implemented using TensorFlow~\cite{abadi2016tensorflow}.
It comprises two sections, an encoder and a decoder.
The encoder compresses the input to a latent dimensional space, whereas the decoder takes the data in the latent layer and decompresses it back to its original size.
The network architecture for the encoder contains two hidden layers, with 800 and 400 neurons respectively, and a latent layer of 200 neurons.
The decoder reverses the structure of the encoder, using 400 and 800 neurons for the two hidden layers, and 1287 neurons for the output layer.
The leaky ReLU~\cite{xu2015empirical} activation function is applied to the output in all hidden and output layers.
The reconstrution loss is defined as the mean squared error between the input and reconstructed values of the dataset.
The logarithm of the reconstruction loss, \logloss, is defined as the anomaly score for each event.
Artificial anomalous events with different characteristics were created and used as input to test the separation power of this architecture.
The default AE outperforms (convolutional) variational AEs.
Furthermore, the selected architecture has shown better separation between anomalous and nonanomalous events than AE architectures with other numbers of neurons.
To form the training and validation datasets, $1\%$ of the collision events are randomly selected after the preselection.
They provide both a sufficient sample size and a good representation of typical collision events.
The chance of finding BSM signal events in these datasets is considered to be negligible.
Even if the training dataset is influenced by BSM physics, such contributions would alter the distribution of the \logloss of the AE and thus the shape of the mass spectra of the background but would not produce discernible bumps~\cite{Chekanov:2021pus}.
Therefore, the search for localized enhancements in the invariant mass distributions should not be affected significantly.
 
The training and validation sets are randomly split using a 7:3 ratio.
The Adam optimizer~\cite{kingma2014adam} with a learning rate of 0.001 is used to train the network by minimizing the \logloss of the training sets.
The batch size is 100 and all events in the training set are reshuffled at the beginning of each epoch.
Training of the AE is monitored via the reconstruction loss of the validation set, and terminated if this value does not decrease within 30 epochs.
To avoid selecting an overtrained model, the training is repeated 50 times with different network initialization and different random splits of the training and validation sets.
The performance is found to be stable and the model that gives the median validation loss among these trials is used.
 
\begin{figure}[h!]
\begin{center}
\includegraphics[width=0.8\textwidth]{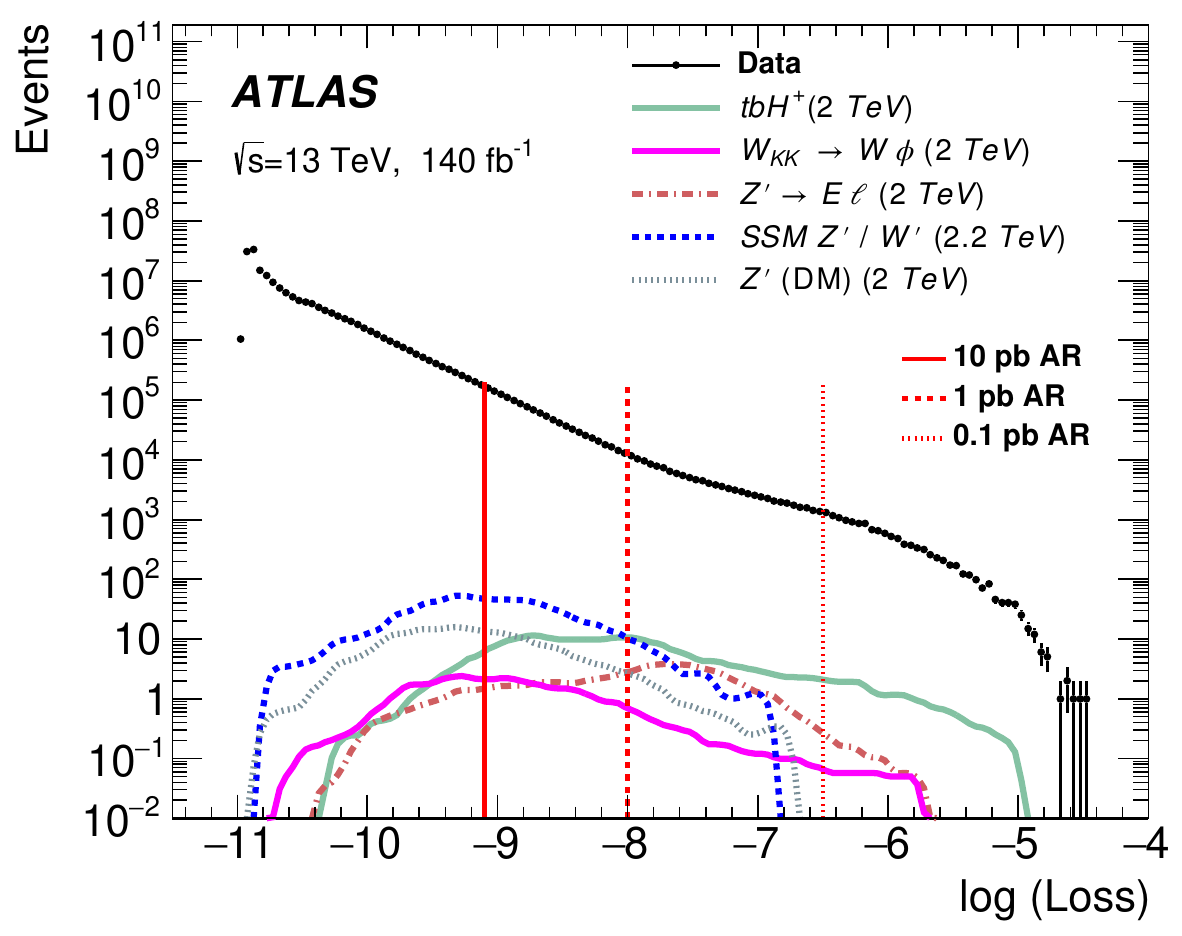}
\end{center}
\caption{
Distributions of the anomaly score from the AE for data and five benchmark BSM models.
Their legends, from top to bottom, are:
\fivemodelwithmass
The BSM predictions represent the expected number of events from \SI{140}{\ifb} of data for heavy particle ($H^+$, $W_\text{KK}$, $Z'$, $W'$ and $Z'$, respectively) masses around 2~\TeV.
The distributions for the BSM models are smoothed to remove fluctuations due to low MC event counts.
The vertical lines indicate the start of the three anomaly regions (ARs).
The labels of the three ARs indicate the visible cross section for hypothetical processes yielding the same number of events as observed in the \SI{140}{\ifb} dataset.
The AE is applied to preselected events without any requirements on invariant mass distributions.
}
\label{fig:BSM_loss}
\end{figure}

The distribution of the anomaly scores of the collision data is shown in \Cref{fig:BSM_loss}.
The data correspond to \SI{140}{\ifb} and include the \SI{1}{\%} of data used in training.
A peak is seen near $\logloss = -10.8$ from the total of \num{166055597} events after preselection.
For comparison, several benchmark BSM models, studied in Refs.~\cite{EXOT-2018-32, EXOT-2020-15}, are overlaid.
These representative models are characterized by the presence of an isolated lepton in the final states.
The models and their parameters are
(1) charged Higgs boson production in association with a top quark, $tbH^{+}$ with $H^+\to t\bar{b}$, with the mass of $H^+$ between 0.4~\TeV\ and 2~\TeV\ with a varying step size~\cite{Degrande:2015vpa}. All top decays are enabled;
(2) a Kaluza--Klein gauge boson, $W_\text{KK}$, with the SM $W$ boson and a radion $\phi$ that decays into two gluons, with the mass of $W_\text{KK}$ between 0.5~\TeV\ and 6~\TeV, and the mass difference of $W_\text{KK}$ and the radion being 0.25~\TeV~\cite{Agashe:2016kfr};
(3) a model with a $Z'$ boson and composite SU(2)$_\text{L}$ fermion doublets that breaks lepton-flavor universality (\enquote{composite lepton}), $Z' \to E\ell$, with the $Z'$ boson mass of 0.5--4~\TeV and various mass hypotheses for the composite lepton $E$ decaying to $Z \ell$ with $Z\to q\bar{q}$~\cite{PhysRevD.98.035010};
(4) the sequential standard model (SSM)  $W'\to W Z' \to \ell\nu q\bar{q}$, with the mass of the $W'$ boson ranging from 0.7~\TeV\ to 6.2~\TeV\ with a varying step size and the mass difference of the $W'$ and $Z'$ bosons being 0.25~\TeV~\cite{Altarelli:1989ff};
(5) a simplified dark-matter (DM) model $Z' \to q\bar{q}$, with an axial-vector mediator $Z'$ boson whose mass ranges between 0.5~\TeV\ and 6~\TeV\ with a varying step size, where one of the quarks radiates a $W$ boson which decays into $\ell\nu$~\cite{Abercrombie:2015wmb}.
As can be seen in \Cref{fig:BSM_loss}, the anomaly scores of the BSM processes tend to be larger than those of the collision events, which are all or mostly produced by SM processes.
The SSM and DM model tend to have characteristics similar to
those of the SM background from the event selection in this analysis, thus yielding lower anomaly scores.
Although only one hypothetical mass is shown for each type of BSM model, it was found that events with larger hypothetical particle masses have larger anomaly scores.
Furthermore, three anomaly regions (ARs) are chosen to maintain sensitivity to different BSM models.
They are defined by $\logloss > -9.1$, $> -8.0$, and $> -6.5$, respectively, as indicated by the vertical lines in \Cref{fig:BSM_loss}.
The labels for the three ARs indicate the visible cross section for hypothetical processes yielding the same number of events as observed in the \SI{140}{\ifb} dataset.
The anomaly score distributions are consistent across the data-taking years, indicating that the AE training is robust against different \pileup conditions and triggering criteria.

The nine invariant mass ($m$) spectra in each anomaly region are examined to search for any localized excesses above
the background hypothesis.
The bin widths chosen for the spectra increase from \SI{16}{\GeV} to \SI{150}{\GeV} over the mass range of 0.3--\SI{8}{\TeV} to reflect the jet energy resolution of the ATLAS detector.
As in the previous ATLAS searches based on the single-lepton requirement~\cite{EXOT-2018-32, EXOT-2020-15}, the following analytic function is used to describe the shape of the smoothly falling background:
\begin{equation}
f(x) = p_1 (1 - x)^{p_2} x^{p_3 + p_4\ln x + p_5 \ln^2 x},
\label{eq:function}
\end{equation}
where $x \equiv m /\sqrt{s}$ and $p_i$\ are free parameters.
 
\begin{figure}[htb]
\begin{center}
\includegraphics[width=0.95\textwidth]{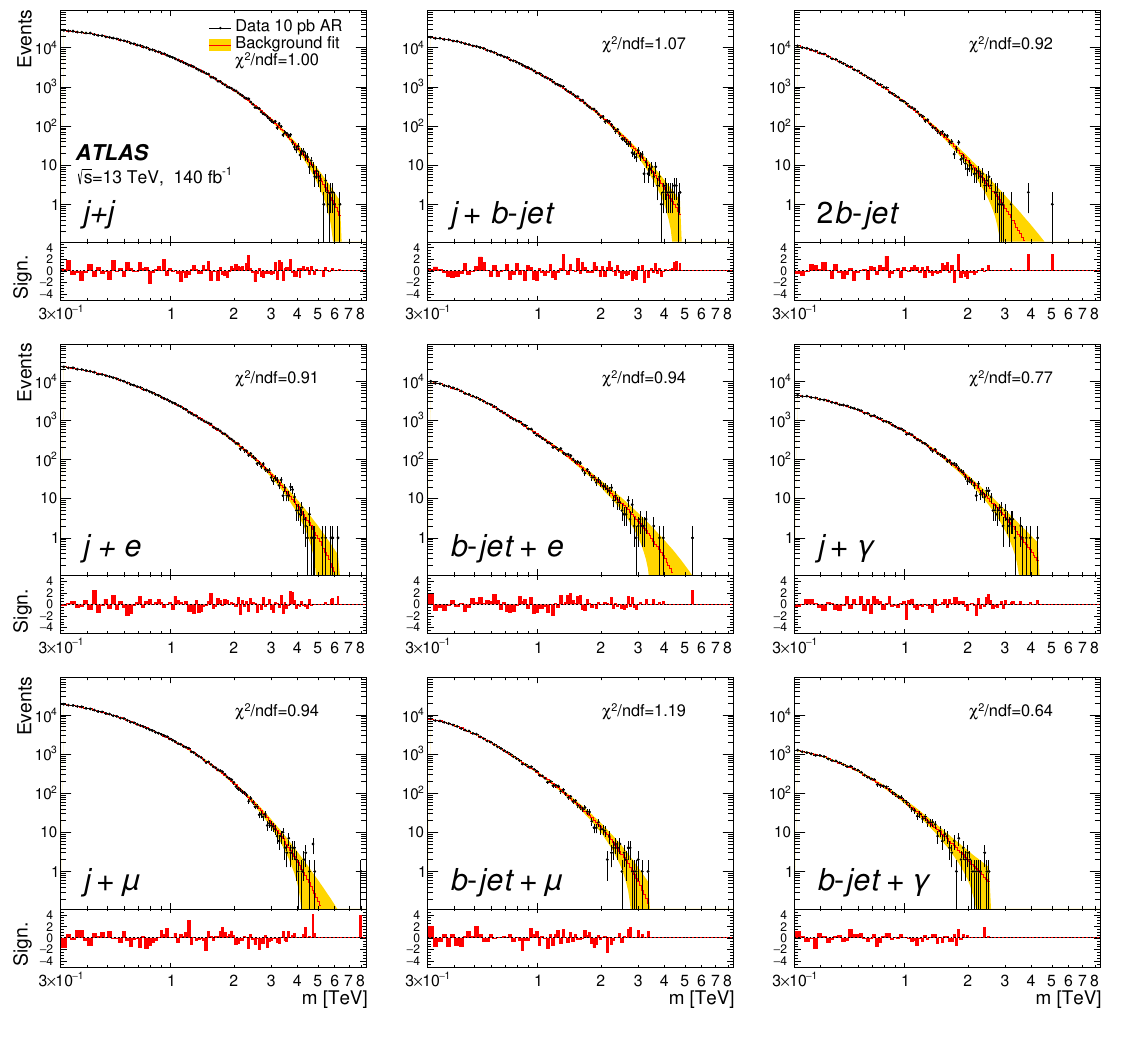}
\end{center}
\vspace{-0.5cm}
\caption{
Invariant mass distributions of jet\;\!+\;\!$Y$ for $\mjy>\SI{0.3}{\TeV}$ in the \SI{10}{\pb} AR along with the fit of \cref{eq:function}.
The fits are represented by the lines, while the associated statistical uncertainties are indicated by the shaded bands.
The lower panels show the bin-by-bin significances of deviations from the fit, calculated as $(d_i - f_i)/\delta_i$, where $d_i$ is the data yield, $f_i$ is the fit value, and $\delta_i$ is the data uncertainty in the $i$-th bin.
}
\label{fig:pub_BH}
\end{figure}

To avoid any possible bias, the analysis was developed without looking at the signal mass distributions in the data.
To verify that \cref{eq:function} can describe the background shape separately for each of the nine invariant mass distributions, fits are performed on the SM background.
This background is estimated from the MC simulation samples, which are composed of $W+$\:\!jets, \ttbar and single-top processes, and a \enquote{loose electron} control sample from data, which represents the QCD multijet events.
Each event in the control sample must contain a leading electron that satisfies the \enquote{loose}, but not the \enquote{tight}, identification criteria~\cite{EGAM-2018-01}, while also passing the anomaly region selections defined by the \SI{1}{\%} data trained autoencoder.
This ensures that the control sample is orthogonal to the signal sample.
The fit results show that \cref{eq:function} can describe the shape of the nine invariant mass distributions in the anomaly regions, with the $\chi^2/$ndf values ranging between 0.7 and 1.9.
The distributions of the fit residuals are consistent with the normal distribution.
More complex fit functions do not improve this description.
These studies do not take into account systematic uncertainties in the SM simulations.
The established functional form of the background hypothesis is then applied to data in the anomaly regions.
Likelihood fits of \cref{eq:function} to the data are performed to determine the free parameters and are shown in \Fig{\ref{fig:pub_BH}} for the \SI{10}{\pb} AR.
The data are well described by the fit, with the $\chi^2/$ndf values ranging between 0.64 and 1.19 for all nine mass spectra.
The lower panels show the fit significances, calculated as $(d_i - f_i)/\delta_i$, where $d_i$ is the data yield, $f_i$ is the fit value, and $\delta_i$ is the data uncertainty in the $i$-th bin.
The distributions of the significances are found to be consistent with normal distributions.
 
Each invariant mass spectrum is analyzed independently using the BumpHunter algorithm~\cite{bumphunter} to look for localized excesses without assumptions about BSM signal shapes.
The uncertainty in the background shape is taken into account by using an alternative function that replaces the term $p_5 \ln^2 x$ with $p_5 / \sqrt{x}$ in \cref{eq:function}.
The algorithm is used to search for statistically significant excesses, taking into account the look-elsewhere effect~\cite{Gross:2010qma} of the full mass spectrum under study and the uncertainty in the background shape.
Among the nine invariant mass distributions in the 10~pb AR, BumpHunter finds the largest excess to be in the $\mjm$ spectrum near 4.8~\TeV, and the second largest to be near \SI{1.2}{\TeV}, as shown in \Cref{fig:pub_BH}.
Their statistical significances when including systematic uncertainties are discussed below.
 
Searches for localized excesses were also performed in the nine invariant mass distributions for the 1~pb and 0.1~pb ARs.
No significant excesses were found.
In particular, the 4.8~\TeV\ region of the $\mjm$ spectra does not show any deviation from the background hypothesis in these two AR regions.
 
\begin{figure}[t!]
\begin{center}
\includegraphics[width=0.8\textwidth]{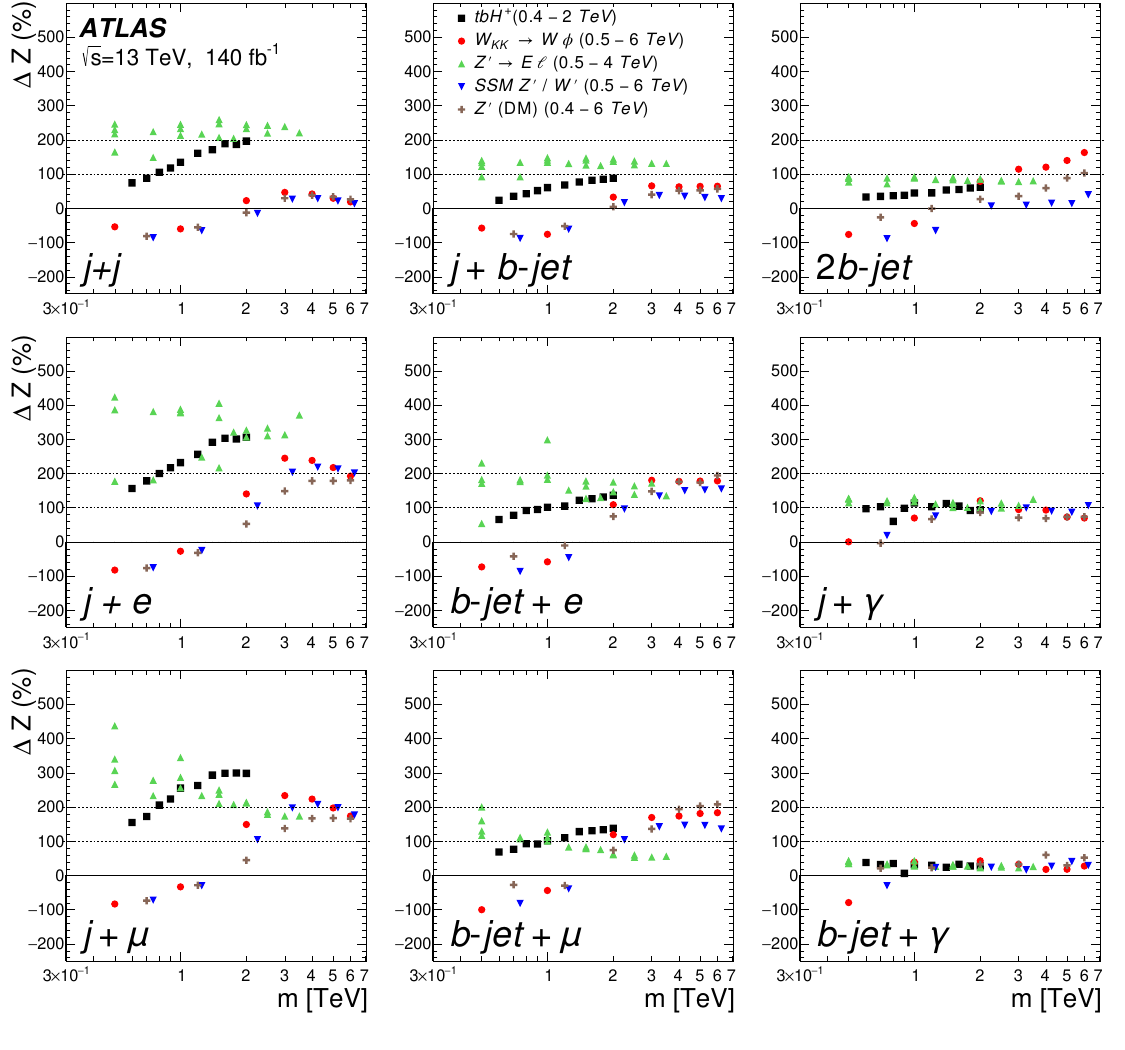}
\end{center}
\vspace{-0.5cm}
\caption{
Values of $\Delta Z$ for the discovery sensitivity, as defined in the text, as a function of the invariant mass $m$. The nine invariant mass distributions are calculated in the \SI{10}{\pb} AR.
Positive percentages indicate improvements in sensitivity. Horizontal dashed lines are drawn at 100\% and 200\% to guide the eye.
The five benchmark BSM models are
\fivemodel
The multiple markers shown for the composite-lepton model at the same invariant mass values correspond to different composite lepton ($E$) masses between 0.25 and 3.5~\TeV. The center positions of the markers are set to the masses of the corresponding heavy particles.
}
\label{fig:pub_BSM_sig}
\end{figure}

The anomaly region selection is expected to improve the discovery potential for BSM models.
\Fig{\ref{fig:pub_BSM_sig}} shows the improvements in discovery sensitivity for the five benchmark BSM models in the \SI{10}{\pb} AR.
The improvement is defined as $\Delta Z = ((Z_{\text{AE}} / Z)- 1) \times 100 \%$, where $Z_{\text{AE}}$ is the discovery sensitivity in the anomaly region and $Z$ is that before any cut on the anomaly score.
The discovery sensitivity is defined as $Z = \sqrt{2 \left(  (s+b)\ln \left( 1+s/b \right) -s \right)}$~\cite{cowan2012}, where $s$ is the number of signal events and $b$ the number of background events calculated from the
fit of \cref{eq:function} to the data.
Both are counted in a $\pm1$-standard-deviation region around the mean value of the reconstructed signal's mass distribution.
The results show that using the AE trained on data improves the discovery sensitivity for most of the benchmark BSM models and mass points.
The BSM hypotheses that do not show improvements in sensitivity correspond to models with low-mass hypothetical particles or models where the number of reconstructed objects and their kinematic-variable values in the RMM are closer to those of SM events.
The 1 and \SI{0.1}{\pb} ARs show less improvements in the sensitivity compared to the \SI{10}{\pb} AR due to significant decrease in statistics.
 
In the absence of any significant resonant signals, upper limits on the cross section times acceptance, efficiency, and branching ratio are set for Gaussian-shaped signals using the \CLs method~\cite{Read:2002hq}.
The effects of systematic uncertainties associated with the luminosity, jet energy scale, jet energy resolution, and object identification and reconstruction are included in the signal.
These object-related uncertainties are derived from the SSM simulated signals.
The uncertainty stemming from the choice of background fit function is taken from the difference of results obtained using \cref{eq:function} and the alternative background shape.
An uncertainty included to account for a fit bias is computed using the spurious-signal method described in Ref.~\cite{HIGG-2016-21}.
This is the dominant source of systematic uncertainty and it weakens the upper limits by approximately $10\%$.  The stochastic fluctuations in the AE training also produce an uncertainty if the AE is retrained.
It can cause a change of the observed limits by 4\% on average.\footnote{This uncertainty is not included in the limits since the anomaly region selection can be treated as a fixed cut when using the trained AE model used in this study.}
 
\begin{figure}[t!]
\begin{center}
\includegraphics[width=0.95\textwidth]{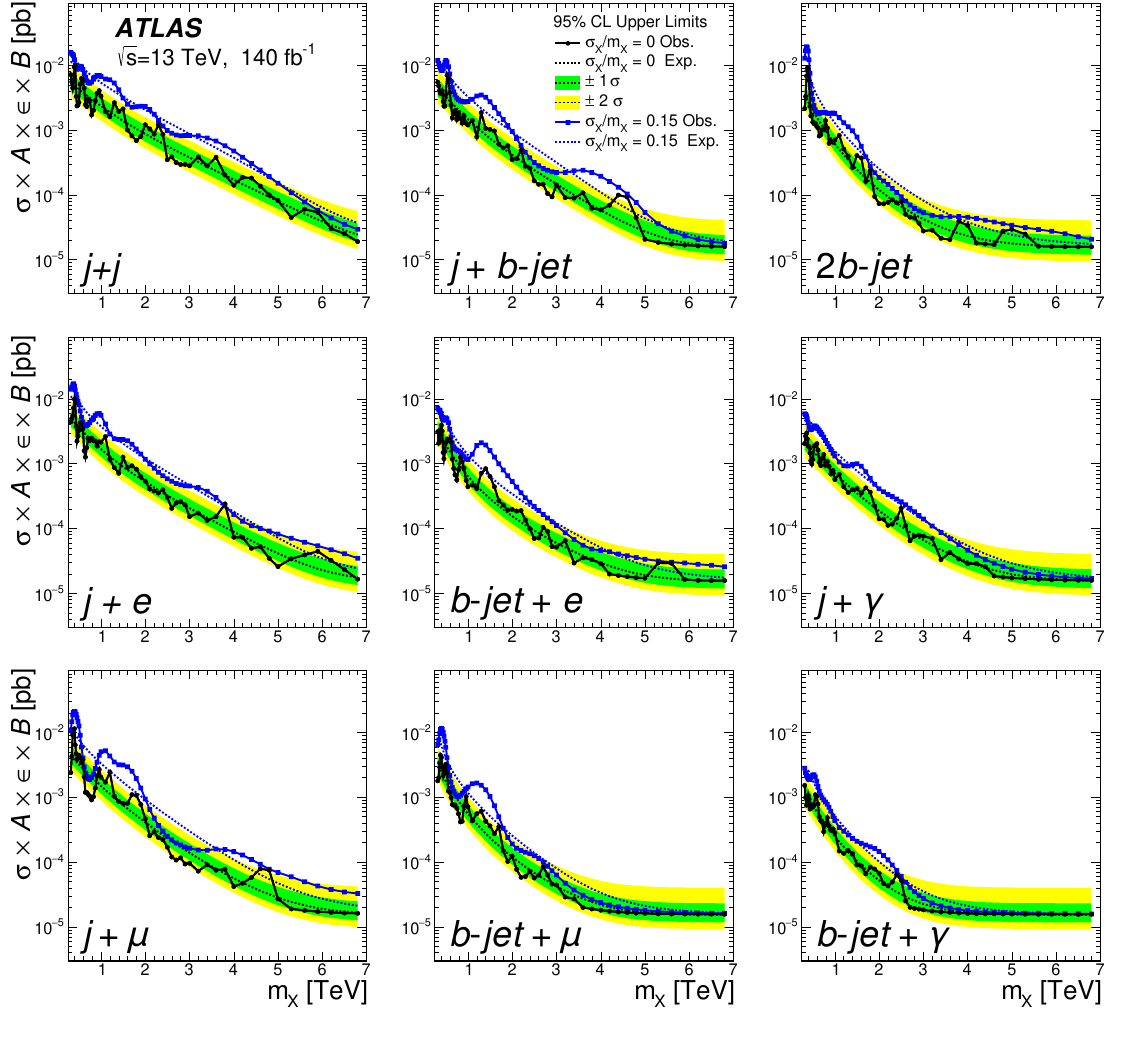}
\end{center}
\vspace{-0.5cm}
\caption{
The 95\% CL upper limits on the cross section times acceptance ($A$), efficiency ($\epsilon$), and branching ratio ($B$) for Gaussian-shaped signals with different signal widths.
The limits are calculated for events with at least one lepton with $\pt^{\ell}>60$~\GeV\ in the \SI{10}{\pb} AR.
Two width hypotheses are shown: $\sigmaX/\mX=0$ and $\sigmaX/\mX=0.15$.
In both cases, the detector resolution for jets is included in the simulation of signal samples.
The $\pm 1\,\sigma$ and $\pm 2\,\sigma$ bands around the expected limit are shown for $\sigmaX/\mX=0$ signals.
Mass points are spaced \SI{5}{\%} apart, relative to the preceding point, starting at \SI{0.3}{\TeV}.
}
\label{fig:gauss_limit}
\end{figure}

Upper limits at 95\% confidence level (CL) on the production cross section of resonant signals with an intrinsic width of \SI{0}{\%} or \SI{15}{\%} of the hypothetical mass in the 10~pb AR are presented in \Fig{\ref{fig:gauss_limit}}.
For a \SI{0}{\%} width, the Gaussian signals have widths consistent  with the experimental resolutions.
The limits are weaker for a \SI{15}{\%} resonance width because the signals are spread over more bins.
The local significance of a signal with a \SI{0}{\%} width for $\mjm = \SI{4.8}{\TeV}$, where the largest excess was detected by BumpHunter, is found to be \Zlimit standard deviations ($\Zlimit\,\sigma$) using the asymptotic formulae~\cite{Cowan:2010js}. A more stringent (\enquote{HighPT}) identification criterion for muons~\cite{MUON-2018-03} removes the event near $\mjm=8$~\TeV, but it does not reduce the significance of the excess at 4.8~\TeV.
The local significance at around $\mjm = \SI{1.2}{\TeV}$ is $\SI{2.8}{\sigma}$.
 
The presented limits are more stringent than the limits on Gaussian signals presented in Ref.~\cite{EXOT-2018-32}, which used the same preselected events.
For example, the limits for $\mjj$ below 1~\TeV\ reported in this analysis show about a factor of 2--3 improvement compared to the results in Ref.~\cite{EXOT-2018-32}.
The search sensitivity is comparable at high masses.
Many BSM models with complex final states are expected to have high AE acceptances.\footnote{The trained AE for evaluating the acceptances of BSM models is available in the HepData repository (record 144864).}
For example, the five benchmark BSM models shown in \Cref{fig:pub_BSM_sig} have an average AE acceptance above 80\% for $\mjj$ masses around 0.5~\TeV\ in the 10~pb AR.
The trained AE rejects about 90\% of the data in the same $\mjj$ region.
This illustrates the potential to set exclusion limits on heavy BSM particles with high AE acceptance.
 
In conclusion, model-independent searches for new phenomena in two-body invariant mass distributions are performed by applying an unsupervised anomaly-detection technique to \lumi~fb$^{-1}$ of $pp$ collision data at $\sqrt{s}=\SI{13}{\TeV}$ recorded by the ATLAS detector at the LHC.
Events are preselected to contain at least one lepton with $\pt^{\ell} > \SI{60}{\GeV}$ and at least one jet with $\pt > \SI{30}{\GeV}$.
An unsupervised machine-learning algorithm based on the AE architecture is trained with a randomly selected \SI{1}{\%} sample of the preselected collision events, without targeting any specific BSM signal.
Three anomaly regions are defined using the reconstruction loss of the AE.
Nine two-body invariant mass spectra, $\mjy$, are studied, where $j$ stands for a light jet or a $b$-jet, and $Y$ is a second light jet or $b$-jet, a lepton, or a photon.
No evidence of significant excesses is observed.
The largest excess reported by BumpHunter is near 4.8~\TeV\ in the $m_{j\mu}$ distribution for the 10~pb AR.
Assuming a resonance with a width of \SI{0}{\%}, the local significance of this excess is $\Zlimit\,\sigma$.
 
The discovery sensitivity shows a large improvement after the anomaly region selection, which is illustrated using several benchmark BSM models in the 10~pb AR.
The analysis sets 95\% CL upper limits
on contributions from generic Gaussian signals
to the studied invariant mass distributions in this AR.
Compared to previous limits without anomaly region selections, the reported model-independent limits have a stronger potential to exclude generic heavy states with complex decay modes.


\FloatBarrier
 
\section*{Acknowledgments}


We thank CERN for the very successful operation of the LHC, as well as the
support staff from our institutions without whom ATLAS could not be
operated efficiently.
 
We acknowledge the support of
ANPCyT, Argentina;
YerPhI, Armenia;
ARC, Australia;
BMWFW and FWF, Austria;
ANAS, Azerbaijan;
CNPq and FAPESP, Brazil;
NSERC, NRC and CFI, Canada;
CERN;
ANID, Chile;
CAS, MOST and NSFC, China;
Minciencias, Colombia;
MEYS CR, Czech Republic;
DNRF and DNSRC, Denmark;
IN2P3-CNRS and CEA-DRF/IRFU, France;
SRNSFG, Georgia;
BMBF, HGF and MPG, Germany;
GSRI, Greece;
RGC and Hong Kong SAR, China;
ISF and Benoziyo Center, Israel;
INFN, Italy;
MEXT and JSPS, Japan;
CNRST, Morocco;
NWO, Netherlands;
RCN, Norway;
MEiN, Poland;
FCT, Portugal;
MNE/IFA, Romania;
MESTD, Serbia;
MSSR, Slovakia;
ARRS and MIZ\v{S}, Slovenia;
DSI/NRF, South Africa;
MICINN, Spain;
SRC and Wallenberg Foundation, Sweden;
SERI, SNSF and Cantons of Bern and Geneva, Switzerland;
MOST, Taiwan;
TENMAK, T\"urkiye;
STFC, United Kingdom;
DOE and NSF, United States of America.
In addition, individual groups and members have received support from
BCKDF, CANARIE, Compute Canada and CRC, Canada;
PRIMUS 21/SCI/017 and UNCE SCI/013, Czech Republic;
COST, ERC, ERDF, Horizon 2020, ICSC-NextGenerationEU and Marie Sk{\l}odowska-Curie Actions, European Union;
Investissements d'Avenir Labex, Investissements d'Avenir Idex and ANR, France;
DFG and AvH Foundation, Germany;
Herakleitos, Thales and Aristeia programmes co-financed by EU-ESF and the Greek NSRF, Greece;
BSF-NSF and MINERVA, Israel;
Norwegian Financial Mechanism 2014-2021, Norway;
NCN and NAWA, Poland;
La Caixa Banking Foundation, CERCA Programme Generalitat de Catalunya and PROMETEO and GenT Programmes Generalitat Valenciana, Spain;
G\"{o}ran Gustafssons Stiftelse, Sweden;
The Royal Society and Leverhulme Trust, United Kingdom.
 
The crucial computing support from all WLCG partners is acknowledged gratefully, in particular from CERN, the ATLAS Tier-1 facilities at TRIUMF (Canada), NDGF (Denmark, Norway, Sweden), CC-IN2P3 (France), KIT/GridKA (Germany), INFN-CNAF (Italy), NL-T1 (Netherlands), PIC (Spain), ASGC (Taiwan), RAL (UK) and BNL (USA), the Tier-2 facilities worldwide and large non-WLCG resource providers. Major contributors of computing resources are listed in Ref.~\cite{ATL-SOFT-PUB-2023-001}.


\clearpage

\printbibliography
 
\clearpage
 
\begin{flushleft}
\hypersetup{urlcolor=black}
{\Large The ATLAS Collaboration}

\bigskip

\AtlasOrcid[0000-0002-6665-4934]{G.~Aad}$^\textrm{\scriptsize 102}$,
\AtlasOrcid[0000-0002-5888-2734]{B.~Abbott}$^\textrm{\scriptsize 120}$,
\AtlasOrcid[0000-0002-1002-1652]{K.~Abeling}$^\textrm{\scriptsize 55}$,
\AtlasOrcid[0000-0001-5763-2760]{N.J.~Abicht}$^\textrm{\scriptsize 49}$,
\AtlasOrcid[0000-0002-8496-9294]{S.H.~Abidi}$^\textrm{\scriptsize 29}$,
\AtlasOrcid[0000-0002-9987-2292]{A.~Aboulhorma}$^\textrm{\scriptsize 35e}$,
\AtlasOrcid[0000-0001-5329-6640]{H.~Abramowicz}$^\textrm{\scriptsize 151}$,
\AtlasOrcid[0000-0002-1599-2896]{H.~Abreu}$^\textrm{\scriptsize 150}$,
\AtlasOrcid[0000-0003-0403-3697]{Y.~Abulaiti}$^\textrm{\scriptsize 117}$,
\AtlasOrcid[0000-0003-0762-7204]{A.C.~Abusleme~Hoffman}$^\textrm{\scriptsize 137a}$
% ...
\bigskip
\\

% ...
$^{29}$Physics Department, Brookhaven National Laboratory, Upton NY; United States of America.\\
% ...
$^{35}$$^{(a)}$Facult\'e des Sciences Ain Chock, R\'eseau Universitaire de Physique des Hautes Energies - Universit\'e Hassan II, Casablanca;$^{(b)}$Facult\'{e} des Sciences, Universit\'{e} Ibn-Tofail, K\'{e}nitra;$^{(c)}$Facult\'e des Sciences Semlalia, Universit\'e Cadi Ayyad, LPHEA-Marrakech;$^{(d)}$LPMR, Facult\'e des Sciences, Universit\'e Mohamed Premier, Oujda;$^{(e)}$Facult\'e des sciences, Universit\'e Mohammed V, Rabat;$^{(f)}$Institute of Applied Physics, Mohammed VI Polytechnic University, Ben Guerir; Morocco.\\
% ...
$^{49}$Fakult\"{a}t Physik , Technische Universit{\"a}t Dortmund, Dortmund; Germany.\\
% ...
$^{55}$II. Physikalisches Institut, Georg-August-Universit\"{a}t G\"ottingen, G\"ottingen; Germany.\\
% ...
$^{102}$CPPM, Aix-Marseille Universit\'e, CNRS/IN2P3, Marseille; France.\\
% ...
$^{117}$Department of Physics, New York University, New York NY; United States of America.\\
% ...
$^{120}$Homer L. Dodge Department of Physics and Astronomy, University of Oklahoma, Norman OK; United States of America.\\
% ...
$^{137}$$^{(a)}$Departamento de F\'isica, Pontificia Universidad Cat\'olica de Chile, Santiago;$^{(b)}$Millennium Institute for Subatomic physics at high energy frontier (SAPHIR), Santiago;$^{(c)}$Instituto de Investigaci\'on Multidisciplinario en Ciencia y Tecnolog\'ia, y Departamento de F\'isica, Universidad de La Serena;$^{(d)}$Universidad Andres Bello, Department of Physics, Santiago;$^{(e)}$Instituto de Alta Investigaci\'on, Universidad de Tarapac\'a, Arica;$^{(f)}$Departamento de F\'isica, Universidad T\'ecnica Federico Santa Mar\'ia, Valpara\'iso; Chile.\\
% ...
$^{150}$Department of Physics, Technion, Israel Institute of Technology, Haifa; Israel.\\
$^{151}$Raymond and Beverly Sackler School of Physics and Astronomy, Tel Aviv University, Tel Aviv; Israel.\\
% ...
$^{a}$ Also Affiliated with an institute covered by a cooperation agreement with CERN.\\
$^{b}$ Also at An-Najah National University, Nablus; Palestine.\\
$^{c}$ Also at Borough of Manhattan Community College, City University of New York, New York NY; United States of America.\\
$^{d}$ Also at Center for High Energy Physics, Peking University; China.\\
$^{e}$ Also at Center for Interdisciplinary Research and Innovation (CIRI-AUTH), Thessaloniki; Greece.\\
$^{f}$ Also at Centro Studi e Ricerche Enrico Fermi; Italy.\\
% ...

\end{flushleft}


\end{document}